\documentclass[12pt,a4paper]{article}
\usepackage{epsf,array,delarray,amsmath,stmaryrd,amssymb}
\usepackage{graphicx}

\newcommand{\sF}{{\cal F}}

\newcommand{\nl}{\vskip 0.05 in \noindent}

\newcommand{\bea}{\begin{eqnarray}}
\newcommand{\eea}{\end{eqnarray}}
\newcommand{\beast}{ \begin{eqnarray*} }
\newcommand{\eeast}{ \end{eqnarray*} }

\newcommand{\be}{\begin{equation}}
\newcommand{\ee}{\end{equation}}

\title{The potential approach in practice}

\author{Tino Kluge\footnote{
Statistical Laboratory, University of Cambridge, 
Wilberforce Road, Cambridge CB3 0WB, UK.
This work was supported by grant number 
GR/D501806 from the Engineering and Physical Sciences 
Research Council}
\\
 L.C.G. Rogers\footnote{
Statistical Laboratory, University of Cambridge, 
Wilberforce Road, Cambridge CB3 0WB, UK}
\\
University of Cambridge}
\begin{document}
\maketitle
\bibliographystyle{plain}

\section{Introduction.}\label{S1}
The basics of what has become known as the {\em potential approach}
to the modelling of interest rate and FX derivatives have been around for
a number of years, and were presented definitively in \cite{potential}.
It is well known that the time-$t$ price $Y_t$ of a contingent claim $Y_T$
to be paid at time $T>t$ can be represented as
\begin{equation}
  Y_t = E[ \, \zeta_T Y_T \,  \vert \, \sF_t \, ] / \zeta_t
\label{pricing}
\end{equation}
where $\zeta$ is the so-called {\em state-price density process}. The
pricing relation (\ref{pricing}) is the foundation of the quantitative
theory of finance, and can be shown to follow very simply from
axioms of linearity, positivity and time-consistency; see, for example,
\cite{origins}.
The state-price density is commonly interpreted as
\begin{equation}
  \zeta_t = \exp\{-\int_0^t r_s \ ds\} Z_t
\label{spd}
\end{equation}
where $r$ is the riskless interest rate, and $Z$ is the likelihood-ratio
martingale which transforms from the reference measure $P$ to the
pricing measure $P^*$. For those coming from an economics 
training, rather than from mathematical finance, the state-price density
would bear the interpretation
\[
   \zeta_t = U'(t,c_t)
\]
as the marginal utility of optimal consumption for an agent in a
general equilibrium.  However one derives the state-price density, the
pricing relation (\ref{pricing}) takes the same form. 

The process $\zeta$ is a positive supermartingale (if the riskless rate
$r$ in (\ref{spd}) is non-negative), and the essential of the potential
approach\footnote{
The name derives from the notion of a potential as a positive 
supermartingale tending to zero in $L^1$.
} is to model the state-price density process {\em directly}. One
way this can be done is to represent the potential as
\begin{equation}
  \zeta_t = E[ A_\infty - A_t \, \vert\, \sF_t \, ]
\label{FH}
\end{equation}
for some integrable increasing process $A$; this is 
in effect what Flesaker \& Hughston do \cite{FH}. However,
for the purposes of calibration, we need a more concrete 
representation, and the initial attempts at calibration always
built the positive supermartingale $\zeta$ in terms of some
Markov process. The seminal paper \cite{potential} studied
a range of diffusion examples, which \cite{RZpot} fitted to 
interest-rate data, with modest success. Taking the underlying
Markov process to be a finite-state chain seemed to fare 
better; see \cite{RYpot}. This is the approach we adopt here,
with some slight variation.

To explain in more detail, we shall suppose that there is a finite-state
Markov chain $X$ taking values in a finite set $I$, with intensity
matrix $Q$, and we shall represent the state-price density process
as 
\begin{equation}
	\zeta_t = f(X_t) \exp( -\int_0^t \alpha(X_s) \; ds )
\label{MCspd}
\end{equation}
for some positive functions\footnote{
The study \cite{RYpot} assumed that $\alpha$ was constant,
a restriction that substantially impairs the quality of fit to data.
} $\alpha, f : I \rightarrow  (0,\infty)$.
In order that the recipe (\ref{MCspd}) defines a supermartingale, 
we expand using It\^o's formula to learn that we shall have to have
\begin{equation}
    (\alpha-Q)f \equiv g \geq 0,
\label{g_def}
\end{equation}
and any non-negative $g$ determines a supermartingale when we take
$f = (\alpha-Q)^{-1} g$.   {\em Therefore a positive supermartingale
(and so a pricing model) is specified by the triple $(Q,\alpha,g)$},
where $Q$ is a Markov chain intensity matrix, and $\alpha$ and
$g$ are non-negative functions.

As is explained in \cite{potential}, we can take (\ref{MCspd})
and (\ref{spd}) to discover that\footnote{
The symbol $\doteq$ signifies that the two sides differ by a local 
martingale.
}
\[
d\zeta_t \doteq -r_t\zeta_t \;dt 
\doteq -(\alpha-Q)f(X_t)e^{ -\int_0^t \alpha(X_s) \; ds } \;dt
\]
and hence that 
\begin{equation}
r_t =  \frac{g(X_t)}{f(X_t)} = \alpha(X_t)-\frac{Qf(X_t)}{f(X_t)}.
\label{r_xp}
\end{equation}
Though this is quite explicit, we have little use of this expression
for the spot rate, as all calculations are handled directly through the 
state-price density parametrized by
\begin{equation}
	\Theta \equiv (Q,\alpha,g).
\label{Theta}
\end{equation}
There are certain redundancies in this parametrization; for example, since the 
row sums of $Q$ must be zero, we only need to record the off-diagonal
entries. Similarly, for any positive $\lambda$, the function $\lambda g$
generates the same model as the function $g$, so we may restrict
attention to the reduced parameter vector\footnote{
Since the entries of $\theta$ are non-negative, in the implementation
we work with $\log \theta$.
}
\[
\theta \equiv ( ( q_{ij})_{i\neq j}, \alpha, (g_i)_{i >1}).
\]
In practice, since the entries of this vector are non-negative, we 
suppose they are positive and work instead with
\begin{equation}
\theta \equiv ( ( \log q_{ij})_{i\neq j}, \log\alpha, (\log g_i)_{i >1}).
\label{theta}
\end{equation}

\section{Calibration: particle filtering}\label{S2}
The potential approach is envisaged as being a framework for 
{\em simultaneously pricing {\bf all} derivatives of interest}, be they
interest rate, credit, equity, FX, hybrid, $\ldots$.  Of course, this is
likely  to be overambitious, but even if we were to regard it as only
being suitable for explaining the prices of interest-rate derivatives
in one currency, we have to recognise that for the pricing of a
general derivative, there will be no closed-form expression, and so we 
will have to resort to some numerical integration. This inevitably means
doing a finite sum, and so the philosophy adopted here\footnote{
... also the philosophy of \cite{RYpot} ..
}
is that we will deal only with Markov processes which take finitely many 
values, that is, {\em finite-state Markov chains}; thus the derivative-pricing
calculation will be an {\em exact} calculation (not an approximation to an 
integral), and this calculation will call on nothing more sophisticated
than (very fast) linear algebra operations.  For example, and in particular, pricing
of an American option will involve the optimal stopping of a finite-state
Markov chain, which is not too difficult to do.  We shall in practice rarely
find any benefit in using more than 10 states for the Markov chain.

A philosophical objection to this approach is that if we work with a 
Markov chain with only (say) 5 states, then at any given time, the model
would only allow a given derivative to have 5 possible values, which 
is hard to believe in the light of the fluctuating behaviour of market
prices of swaptions, for example.  We shall get round this objection, and 
address the concrete questions of calibration, by using a {\em particle
filtering} point of view.

Particle filtering is a computational Bayesian methodology for filtering
the state of a hidden (discrete-time) Markov process $(x_t)$ from
observations $(Y_t)$. We suppose that the Markov process has transition
density $p(x'|x)$, and that the likelihood\footnote{
This is not the most general form of the particle-filtering setup, because it
may be that some part of the Markovian state $x$ is actually observable;
in the applications of interest here, however, this will not happen.
} of the observation $y$ given the (hidden) state $x$ is $f(y|x)$.
The posterior likelihood $\pi_t$ at time $t$,  is approximated by a 
finite collection of point masses:
\begin{equation}
  \pi_t \simeq \sum_{i=1}^N w_t^i \delta_{x_t^i}.
\label{PF1}
\end{equation}
The updating step from one time $t$ to the next $t'$ is achieved by moving
each particle $x_t^i$ to a randomly-chosen position $x_{t'}^i$
according to a density $q(\cdot| x_t^i, y_{t'})$ which 
 may depend on the next observation. The 
simplest proposal density $q$ would simply be the transition density
$p$, but we have\footnote{
The point is that if we simply pick $x_{t'}^i$ according to $p(\cdot|x_t^i)$,
then all of the $x_{t'}^i$ may be massively inconsistent with the new observation
$y_{t'}$.
} to be able to tilt the proposal density towards the new data point.  Having
chosen the new points, we re-weight them by
\begin{equation}
   w_{t+1}^i \propto  \frac{w_t^i  \, p(x_{t'}^i | x_t^i)}
				{q(x_{t'}^i| x_t^i, y_{t'})} 
		f(y_{t'} | x_{t'}^i),
\label{PF2}
\end{equation}
where the constant of proportionality is such as to make the weights sum to 
one.

The particle-filtering methodology is a simple generic method, universally
applicable, and as such, dependent on careful tailoring to work well on 
any given example; any special features must be understood and exploited 
if the methodology is to succeed. Here are some of the particular features 
of our situation.

\nl
(1) {\em The Markovian state is
$x = (\xi,\theta)$, where $\xi$ is a finite-state Markov chain. }
Since parameters do not change, if we just update by the transition
density we will never change the set of possible values of the
parameter $\theta$, so the posterior can never move to the
true value. Clearly this is unsatisfactory, so we will 
introduce a (small) `shake' of the parameters at each step, 
shifting $\theta$ to $\theta'$ according to transition density $\kappa(\theta,
\theta')$. In terms of the theory, this corresponds to approximating
the posterior $\pi_t$ not by (\ref{PF1}) but by
\begin{equation}
  \pi_t \simeq \sum_{i=1}^N w_t^i \; \delta_{\xi_t^i}\otimes 
\kappa(\theta_t^i,\cdot).
\label{PF3}
\end{equation}
Operationally, we are introducing a simulated annealing step, and
we continue to denote by $p(\cdot|\cdot)$ the transition density of
the $x_t$, although this now incorporates the possible movement of
the $\theta$-values.
We tried various forms of $\kappa$; gaussian, mutivariate-$t$, 
or Laplace\footnote{
The components of the Laplace distribution are independent and
symmetric, and their absolute values are exponentially distributed.
}.

However, simply shaking the parameters $\theta$ and trusting to
the particle-filtering algorithm is not satisfactory in this application, 
since the dimension of the space is typically too large. For the success
of the method, it is crucial that updates of the particles are importance-sampled
as we now describe.

The new observation $y_t$ is a vector of asset prices\footnote{
Some of the data are swap rates.
} which we imagine are modelled as a function $\eta(x)$ of some
unknown $x=(\xi,\theta)$, plus some noise\footnote{
Since we are free to 
permute the labelling, we shall suppose at this point that $\xi=1$
for the purposes of calculating prices.
}. We suppose that the likelihood of $y_t$ given $x$ is of the form
\[
	f(y|x) = \varphi_Y(\log y-\log\eta(x))
\]
for some suitable density $\varphi_Y$  concentrated around zero, and initially
{\em we simply seek out the maximum-likelihood estimator
of $\theta$:}
\begin{equation}
    \theta^* = \hbox{\rm arg max} \;  f(y_t|(1,\theta)).
\end{equation}
This step is quite computationally intensive (we use a combination
of gradient search methods, and simulated annealing), but
seems to be  unavoidable.
Notice that the $\theta^*$ identified does not depend in any way on the
previous particle population. However, we use the MLE $\theta^*$ to 
pull the proposed values of $\theta$ into a plausible region, and then
we reweight them. In more detail, for a given particle $x^i_{t-1}
=(\xi_{t-1}^i, \theta_{t-1}^i)$ we first generate a new $\theta$-value 
$\theta_t^i$
according to a density $\varphi_\theta(\cdot - \theta^*)$,
and move the state $\xi$ according to the dynamics implied by the
newly-chosen $\theta$-value, creating the new particle $x_t^i$. The
new weight attached to this particle is proportional to
\[
	w_{t-1}^i \; \frac{p(x_t^i|x_{t-1}^i) f(y_t|x_t^i)}{
   \varphi_\theta(\theta^i_t-\theta^*)}.
\]

\nl
(2) {\em The model price is the population average of the 
individual particle prices.} This is a straightforward application of 
Bayesian ideas; given some derivative, and a posterior 
$\pi_t$ given in the form (\ref{PF1}), we calculate  the particle
price $\eta(x_t^i)$ of the derivative for each particle, and then take
as the {\em model price} the average
\[
	\sum_{i=1}^N w_t^i \, \eta(x_t^i).
\]
Contrast this with what would happen were we to try to follow
some classical maximum-likelihood approach. At each time, we would
calculate some MLE of the parameters of the problem, and then we 
are faced with the philosophical difficulty that if we believe that the
current MLE is the truth, then the price for any given asset can only
take $n$ values (where $n$ is the number of states of the chain.)
The Bayesian approach glides over this problem; at any time, the 
price of a derivative is some posterior average of the prices which
would arise under different models, and different values of the state
of the chain, so there is no problem of there being conceptually only
a small number of possible prices.  Moreover, the Bayesian 
particle-filtering approach gives us at any stage a posterior distribution
for the price of {\em any} derivative, and could be used provide 
confidence intervals for the price.  This could have important 
practical applications; industry calibrations typically insist on 
exact matching of `the' market prices of the calibration
instruments\footnote{
... typically  collected at different times and  exchanges ....},
and this leads to some very silly modelling - in fact, fitting, not
modelling.  But an approach which computes confidence intervals
for prices reflects uncertainty in the outputs of the model, 
driven by the uncertainty in the inputs to the model, and would 
allow a successful calibration to be defined in terms of `the'
calibration prices lying inside some confidence interval.

Implementing the particle filtering algorithm requires some 
care. There is the generic problem of {\em impoverishment}, where
after a time all but a few of the particles have almost zero
weight, so that evolving those particles is a waste of effort.
We deal with this problem by the usual {\em resampling}
technique.
Next there is the problem of choosing the `shake', expressed through
the transition density $\kappa$. What should be the distribution of the
shake, how should it be scaled\footnote{We compare the parameter changes
of a series of ML estimates and choose a similar distribution for
the parameter shakes. Here we see a standard deviation of around $0.05$ in log
parameter space.}?
But the most important 
problem is 
finding the optimal parameters $\theta^*$.
As a numerical method which is capable of
finding a global minimum we use a combination
of gradient methods to converge to local minima, and
simulated annealing steps to try to avoid non-global minima.
The curse of dimensionality makes it very hard to obtain
nearly optimal solutions for more than 10 markov states
(10 markov states then the space of parameters $\theta$ is of
dimension $109$). Some ideas help to reduce the dimensionality.

Based on many simulations we find that restricting the $Q$ matrix
to be a nearest-neighbour Markov chain on a circular state-space
does not necessarily reduce the quality of fit. In fact, MLEs obtained
by simulated annealing with a full $Q$ matrix and nearest-neighbour
chain give fits of  similar quality. The dimensionality reduction
in $\theta=(Q,\alpha,g)$ obviously helps the numerical optimisation
algorithm. However, if $Q$ was restricted to a nearest-neighbour
Markov chain on a circle that was only allowed to travel in one
direction, then the quality of fit is much worse.


\section{The data.}\label{S3}
The data we worked with was daily data for the period 23rd April 2003 
until 1st January 2007, and consisted of
\begin{itemize}
 \item LIBOR rates: 1m, 3m, 6m, 12m
 \item Swap rates: 2y, 3y, 5y, 7y, 10y
 \item Cap prices: 1y, 3y, 5y, 7y, 10y (at-the-money strike)
 \item Swaption prices: 6m, 1y, 2y, 3y, 5y into 2y, 3y, 5y, 7y, 10y (at-the-money strike)
\end{itemize}
in four currencies (USD, EUR, GBP, JPY), along with FX forwards into USD
of the other three currencies, looking ahead 1m, 3m, 6m, 1y.
For swap, swaptions and caps, payments are quarterly.  The data
were quite clean, and represented an excellent source to work from.

\section{Results of the fitting.}\label{S4}
\begin{figure}
  \centering
  \includegraphics[width=1.0\textwidth]{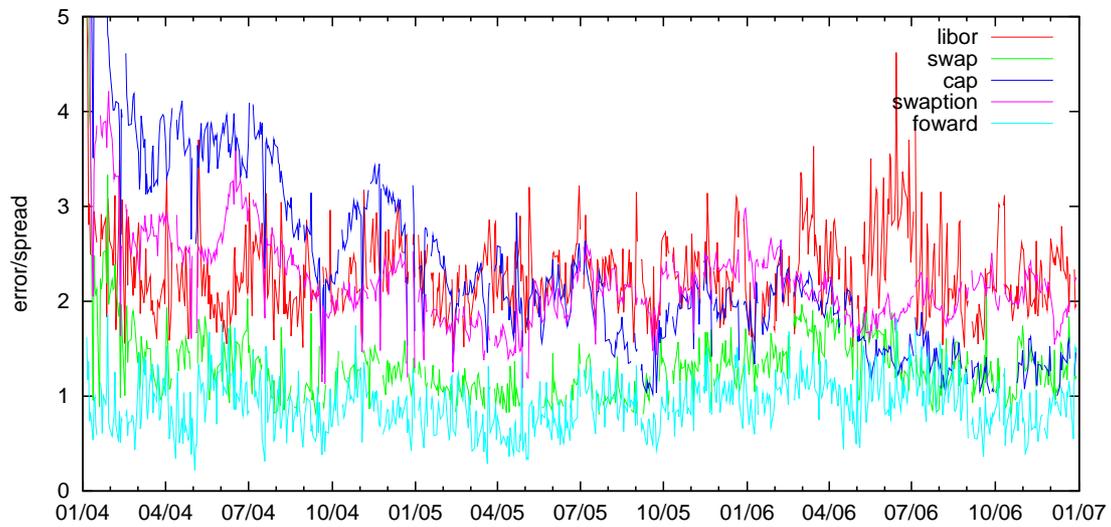}
  \caption{Average absolute errors in spreads.}
  \label{errors_bp}
\end{figure}

We present various plots to summarise the quality of the fits. In the
first, Figure \ref{errors_bp}, we show how the average absolute
 errors (measured in spreads\footnote{
We took the spreads in  ... to be ...
})  vary across time. The averages are split according to the types of
instrument.  It is interesting
to note that  for FX forwards and swaps the average error is typically
no bigger than 1.5 spreads, and for caps, swaptions, and Libor rates
the errors are of the typical order of 2.5 spreads.

\begin{figure}
  \centering
  \includegraphics[width=0.95\textwidth]{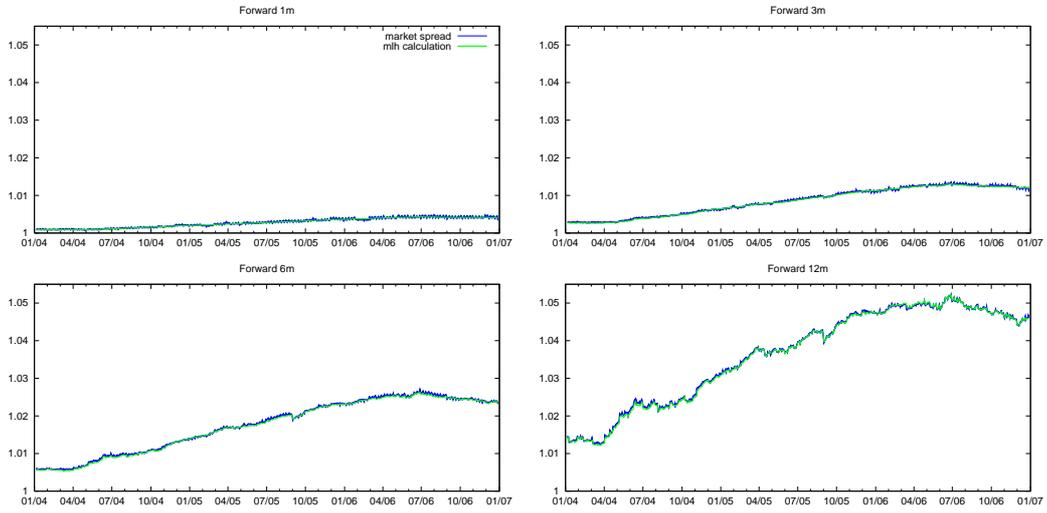}
  \caption{Forward JPY rates.}
  \label{prices_forward_jpy}
\end{figure}

\begin{figure}
  \centering
  \includegraphics[width=0.95\textwidth]{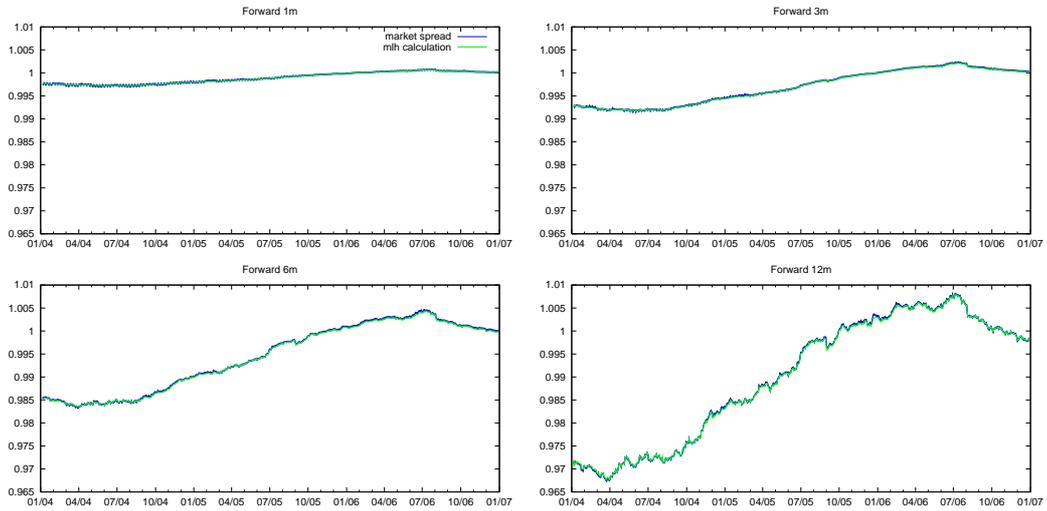}
  \caption{Forward GBP rates.}
  \label{prices_forward_gbp}
\end{figure}

\begin{figure}
  \centering
  \includegraphics[width=0.95\textwidth]{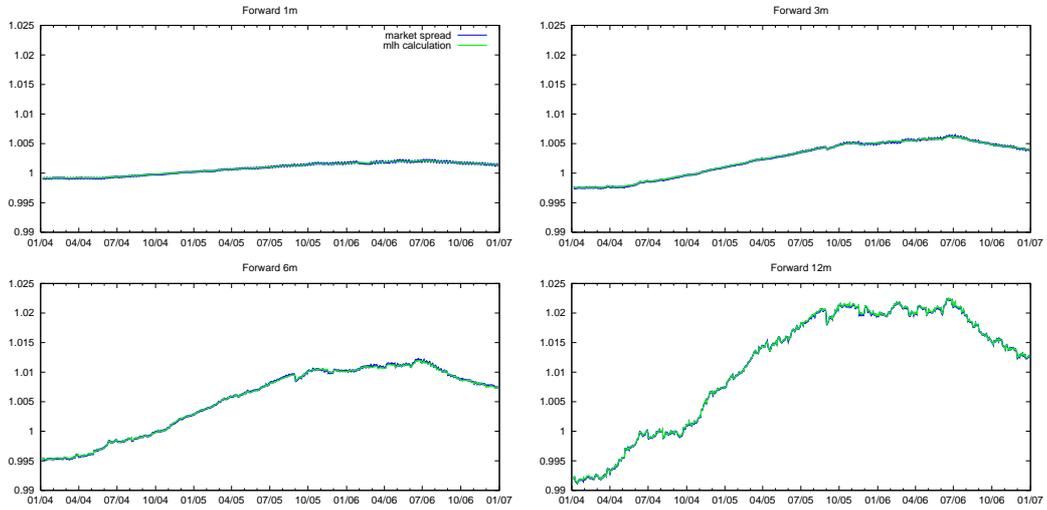}
  \caption{Forward EUR rates.}
  \label{prices_forward_eur}
\end{figure}

\begin{figure}
  \centering
  \includegraphics[width=0.95\textwidth, height=0.399\textheight]{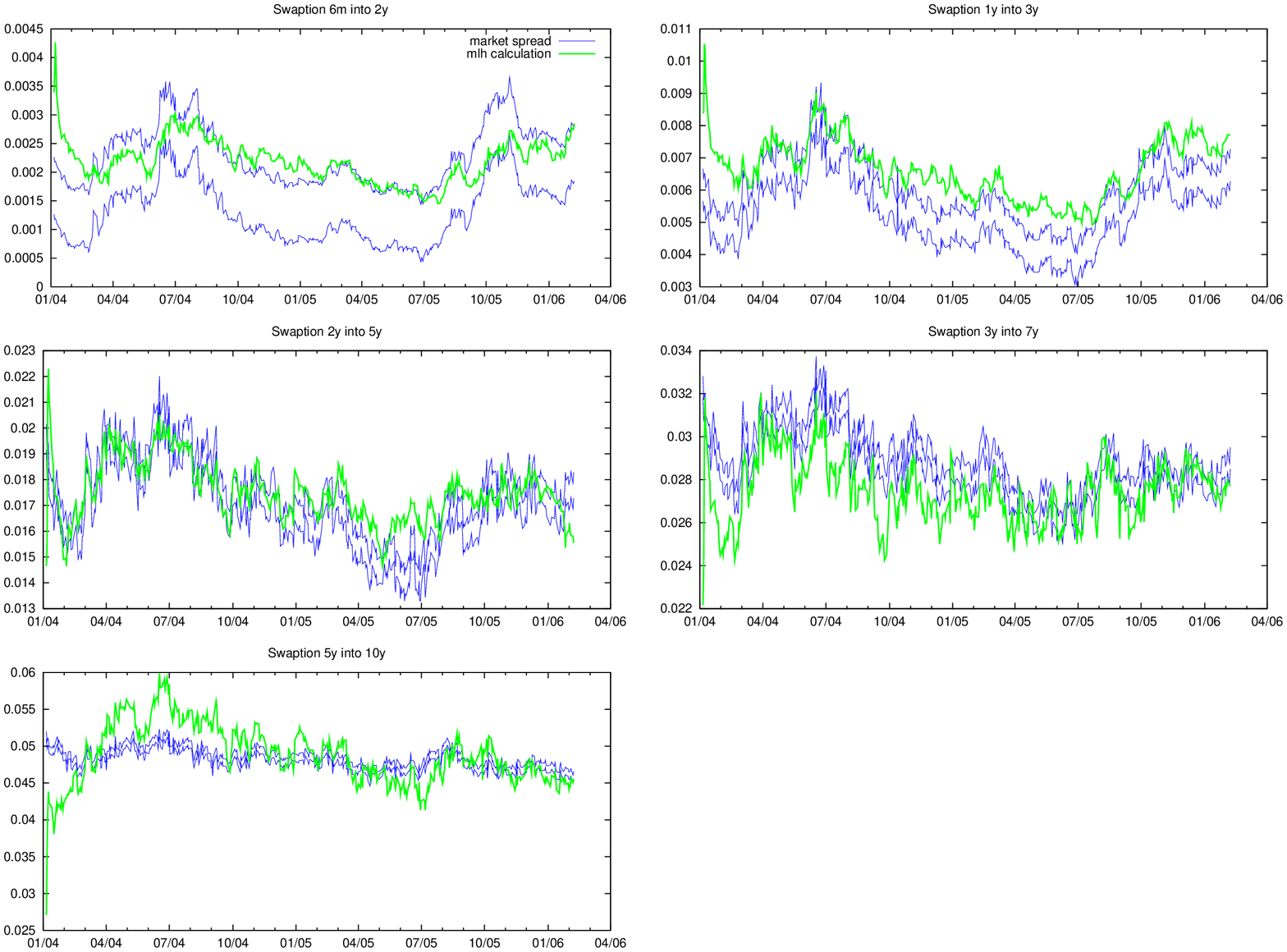}
  \caption{JPY swaption prices.}
  \label{prices_swaption_jpy}
\end{figure}

\begin{figure}
  \centering
  \includegraphics[width=0.95\textwidth, height=0.399\textheight]{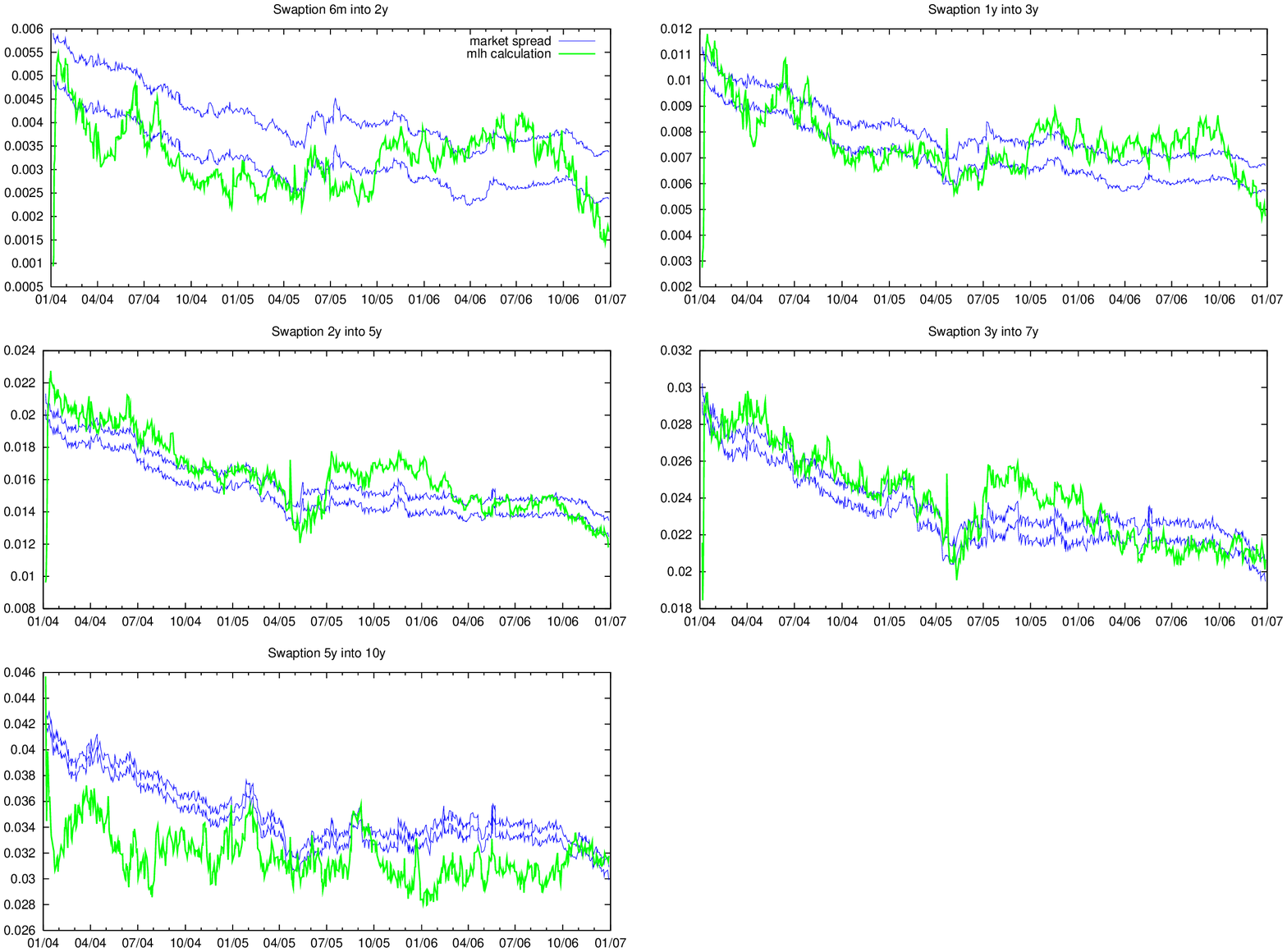}
  \caption{GBP swaption prices.}
  \label{prices_swaption_gbp}
\end{figure}

\begin{figure}
  \centering
  \includegraphics[width=0.95\textwidth, height=0.399\textheight]{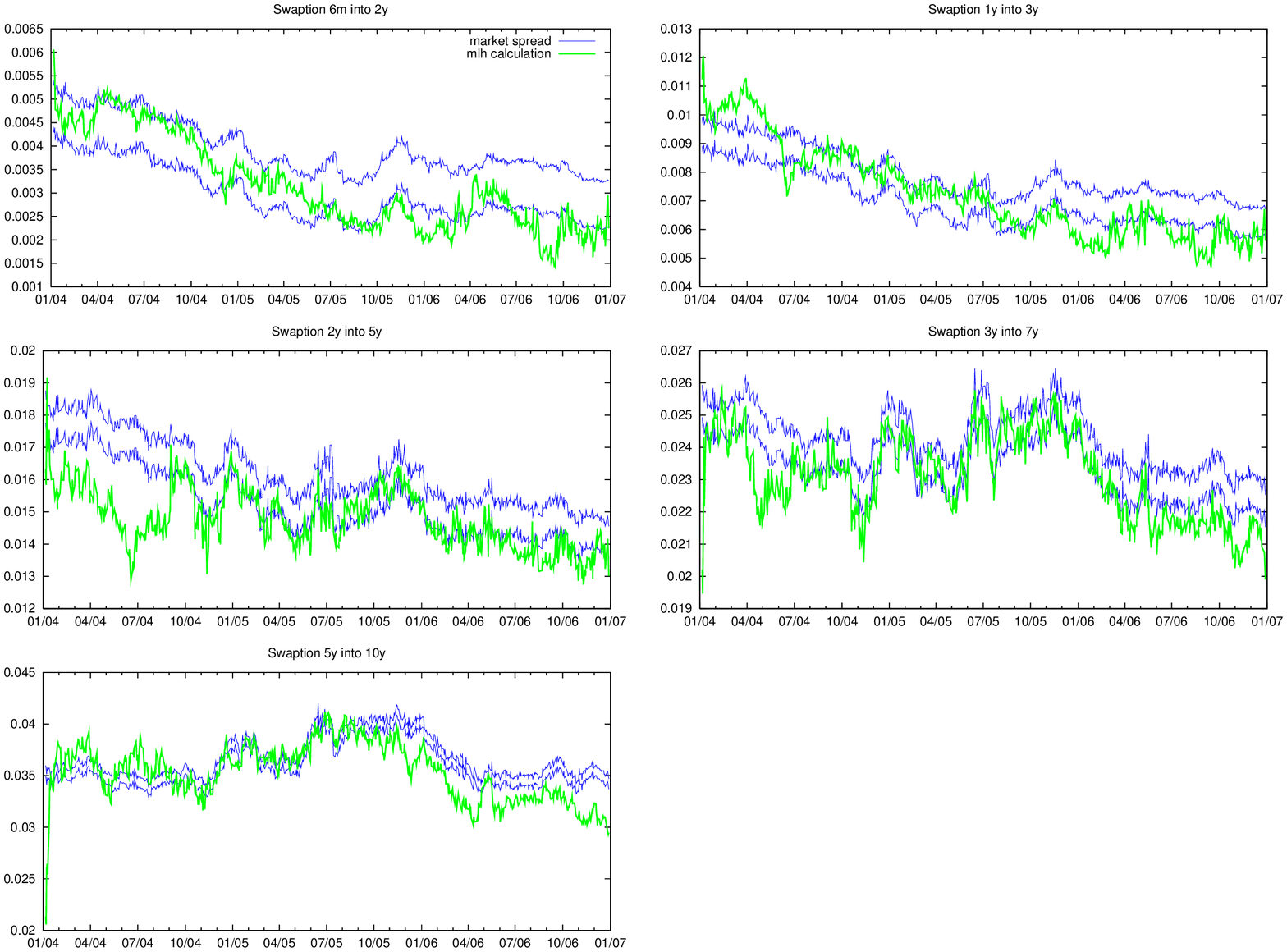}
  \caption{EUR swaption prices.}
  \label{prices_swaption_eur}
\end{figure}

\begin{figure}
  \centering
  \includegraphics[width=0.95\textwidth, height=0.399\textheight]{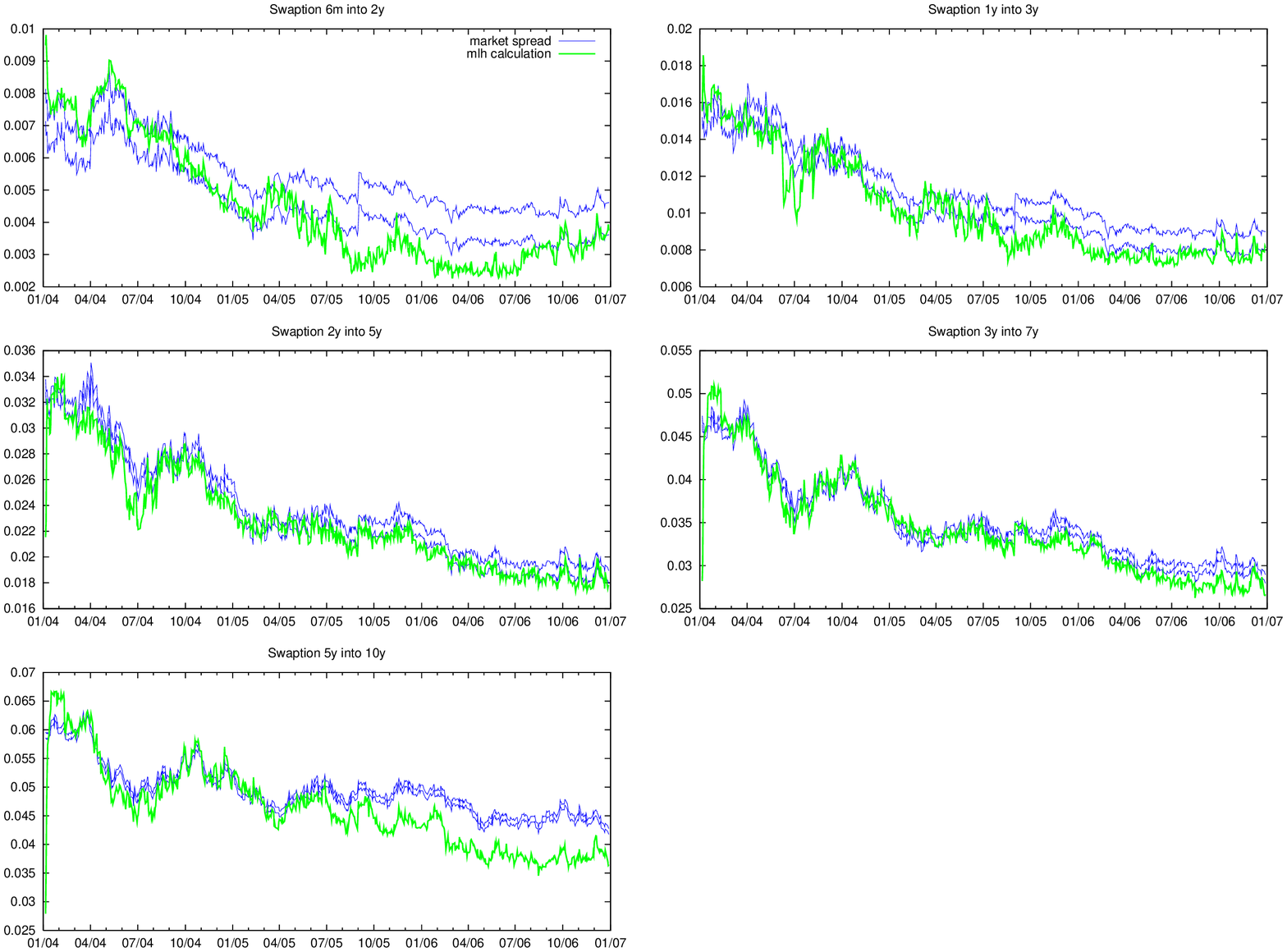}
  \caption{USD swaption prices.}
  \label{prices_swaption_usd}
\end{figure}

\begin{figure}
  \centering
  \includegraphics[width=0.95\textwidth, height=0.399\textheight]{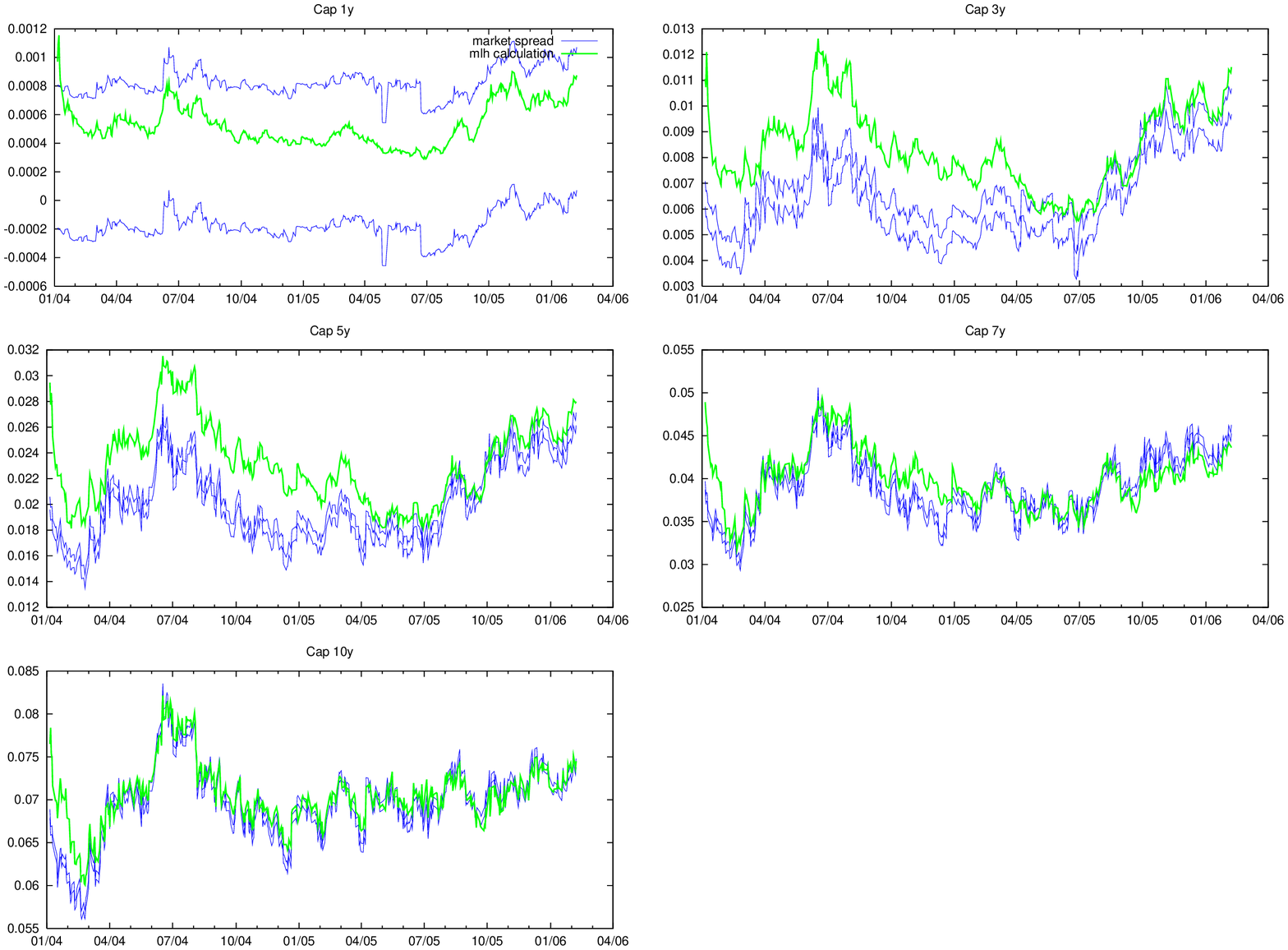}
  \caption{JPY cap prices.}
  \label{prices_cap_jpy}
\end{figure}

\begin{figure}
  \centering
  \includegraphics[width=0.95\textwidth, height=0.399\textheight]{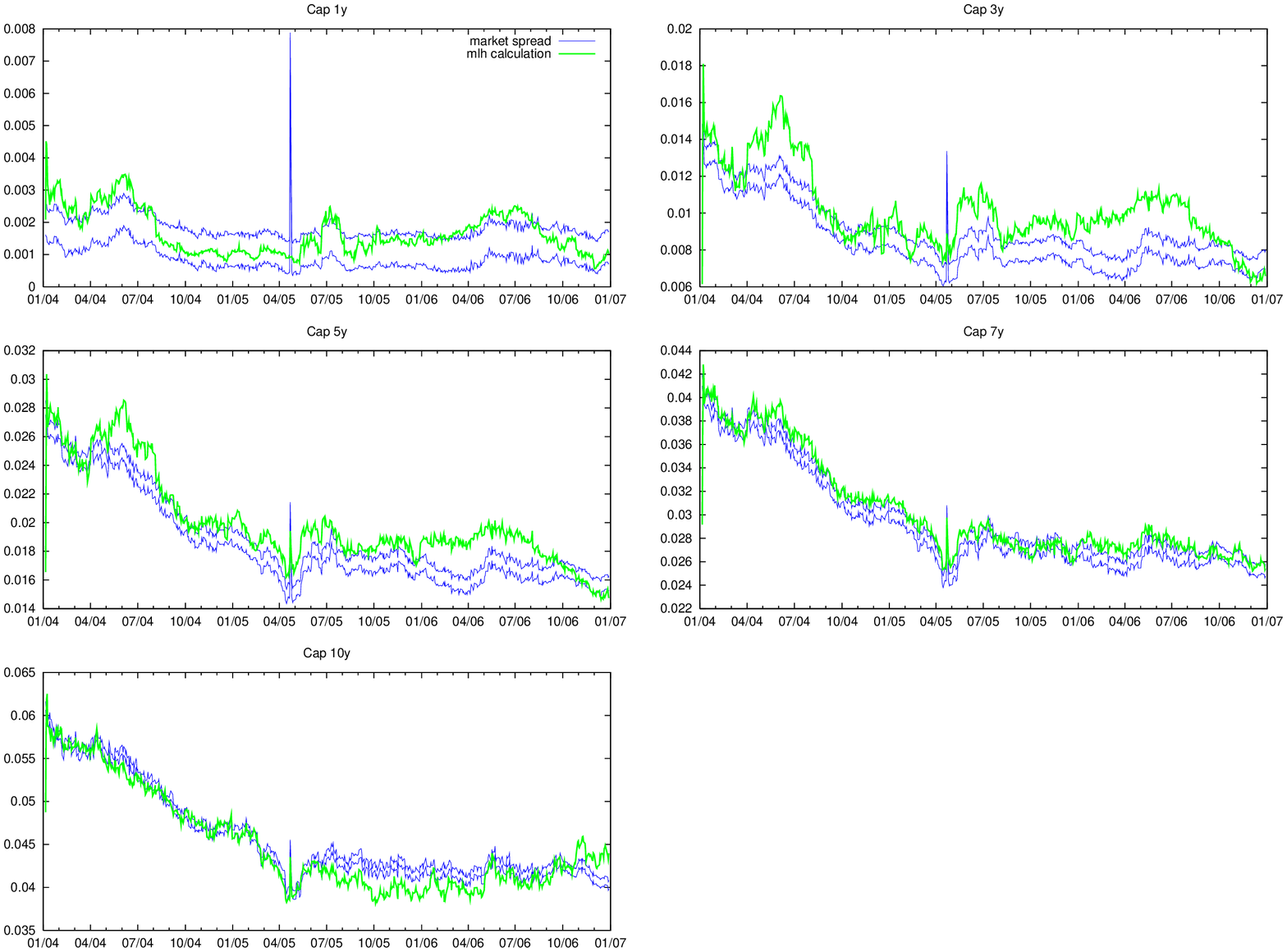}
  \caption{GBP cap prices.}
  \label{prices_cap_gbp}
\end{figure}

\begin{figure}
  \centering
  \includegraphics[width=0.95\textwidth, height=0.399\textheight]{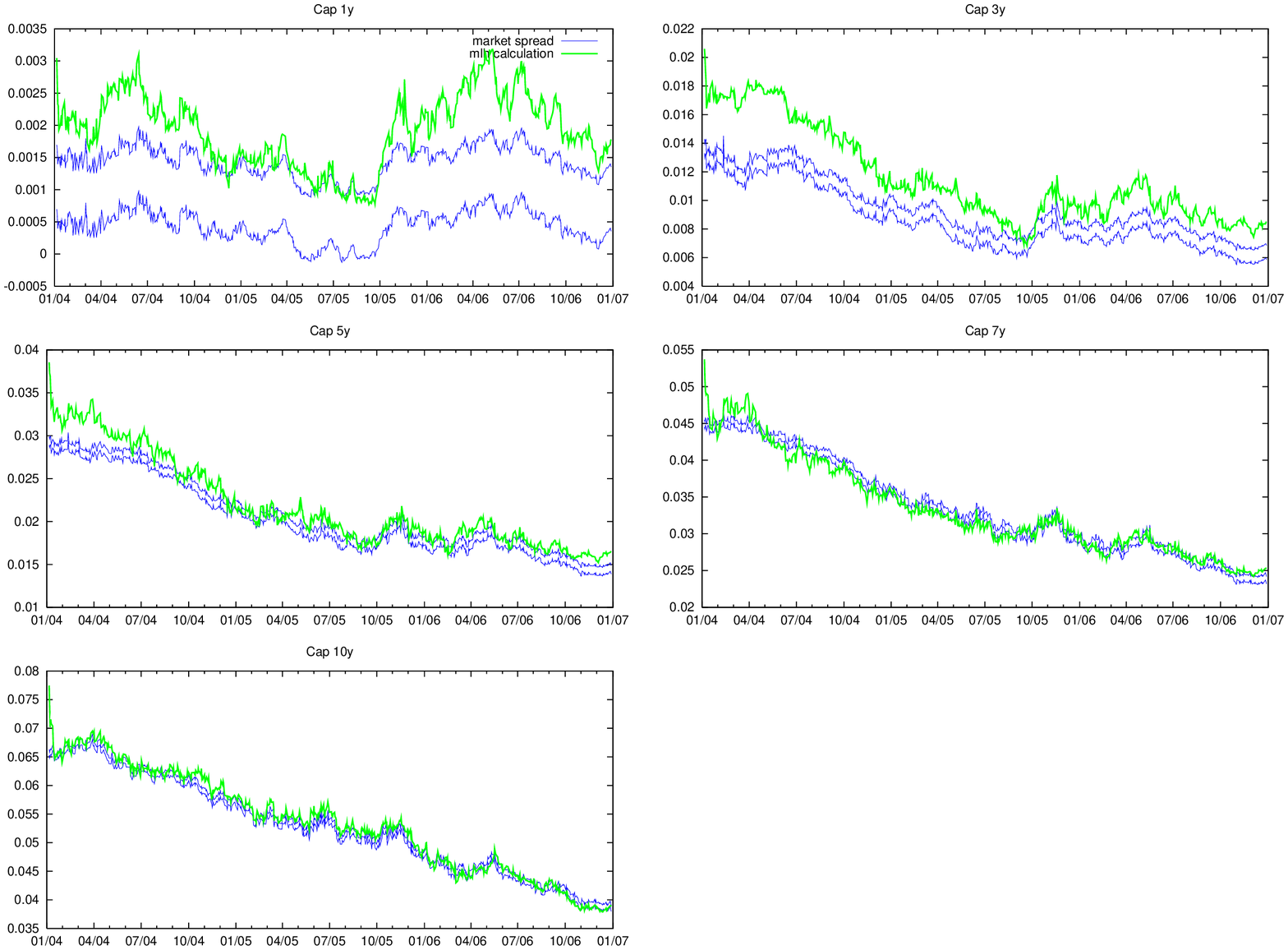}
  \caption{EUR cap prices.}
  \label{prices_cap_eur}
\end{figure}

\begin{figure}
  \centering
  \includegraphics[width=0.95\textwidth, height=0.399\textheight]{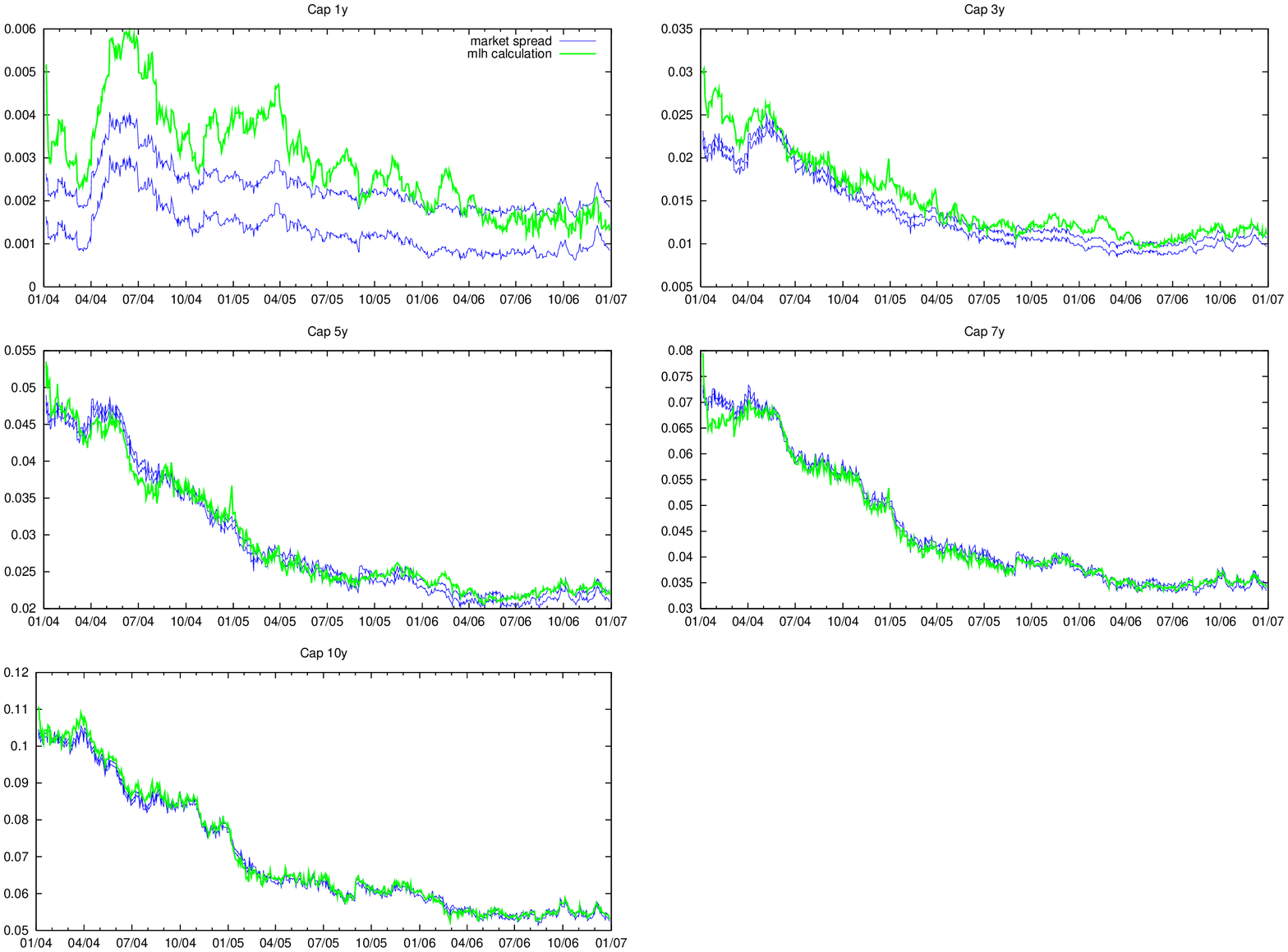}
  \caption{USD cap prices.}
  \label{prices_cap_usd}
\end{figure}

\begin{figure}
  \centering
  \includegraphics[width=0.95\textwidth, height=0.399\textheight]{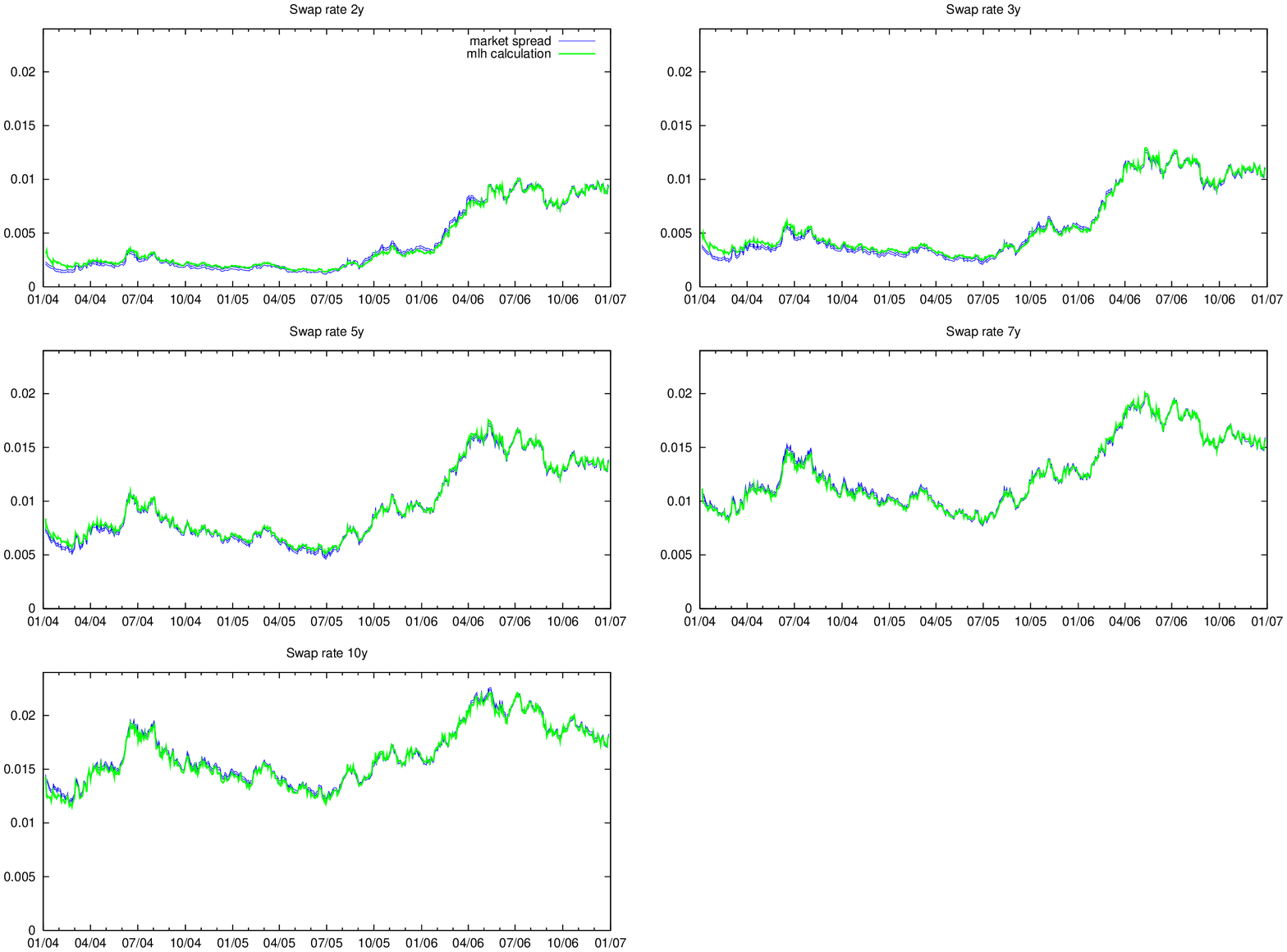}
  \caption{JPY swap rates.}
  \label{prices_swap_jpy}
\end{figure}

\begin{figure}
  \centering
  \includegraphics[width=0.95\textwidth, height=0.399\textheight]{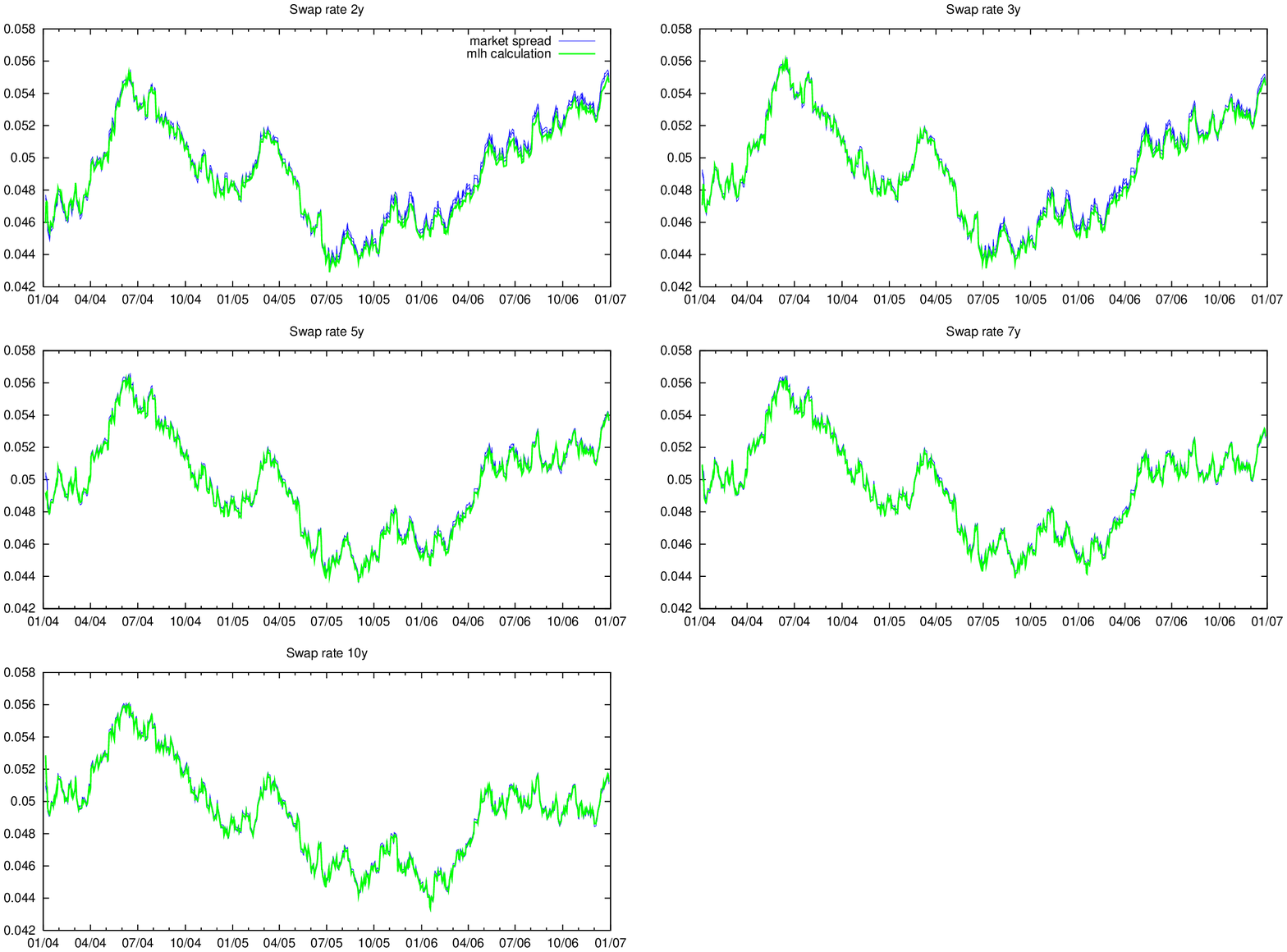}
  \caption{GBP swap rates.}
  \label{prices_swap_gbp}
\end{figure}

\begin{figure}
  \centering
  \includegraphics[width=0.95\textwidth, height=0.399\textheight]{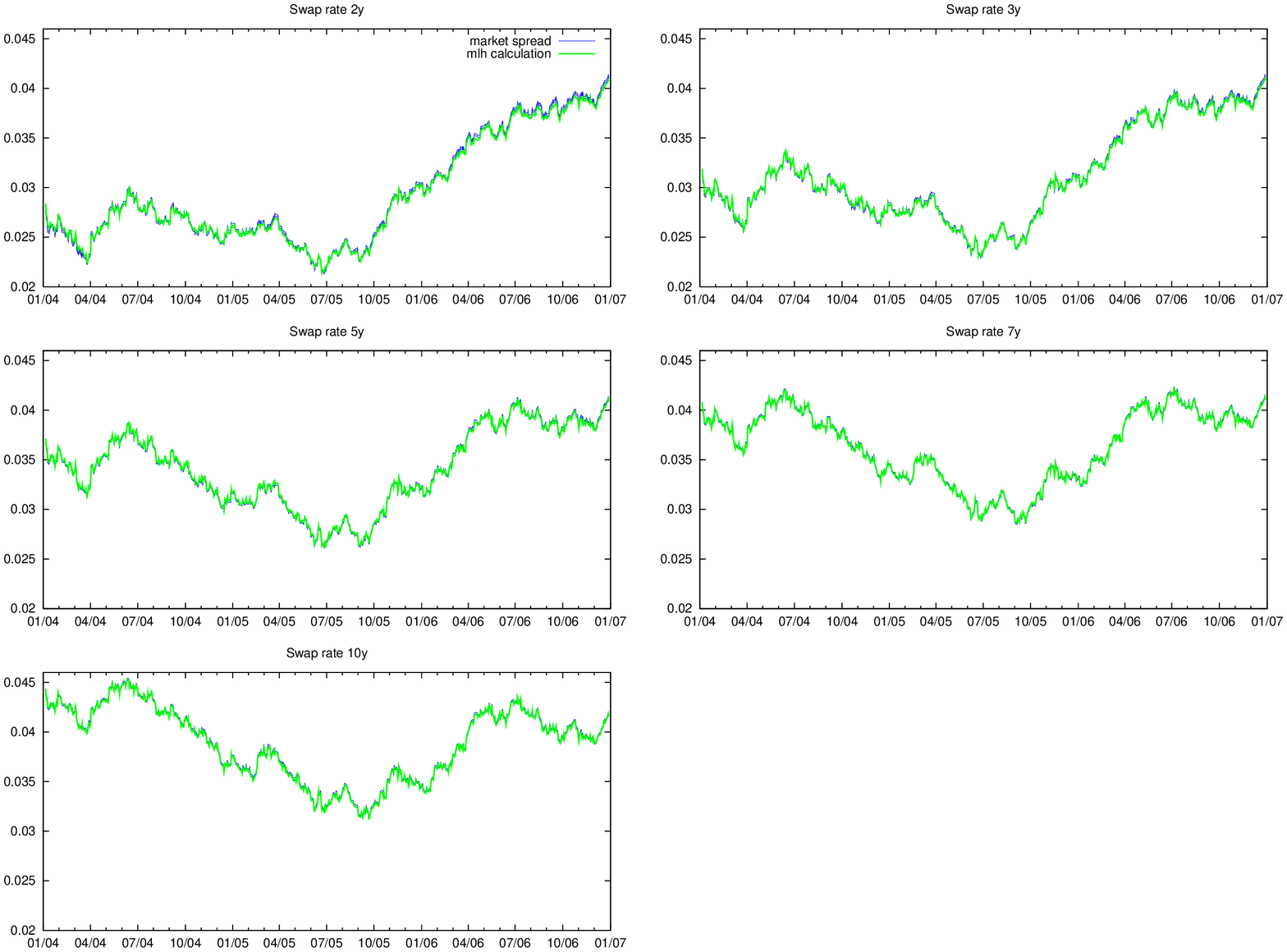}
  \caption{EUR swap rates.}
  \label{prices_swap_eur}
\end{figure}

\begin{figure}
  \centering
  \includegraphics[width=0.95\textwidth, height=0.399\textheight]{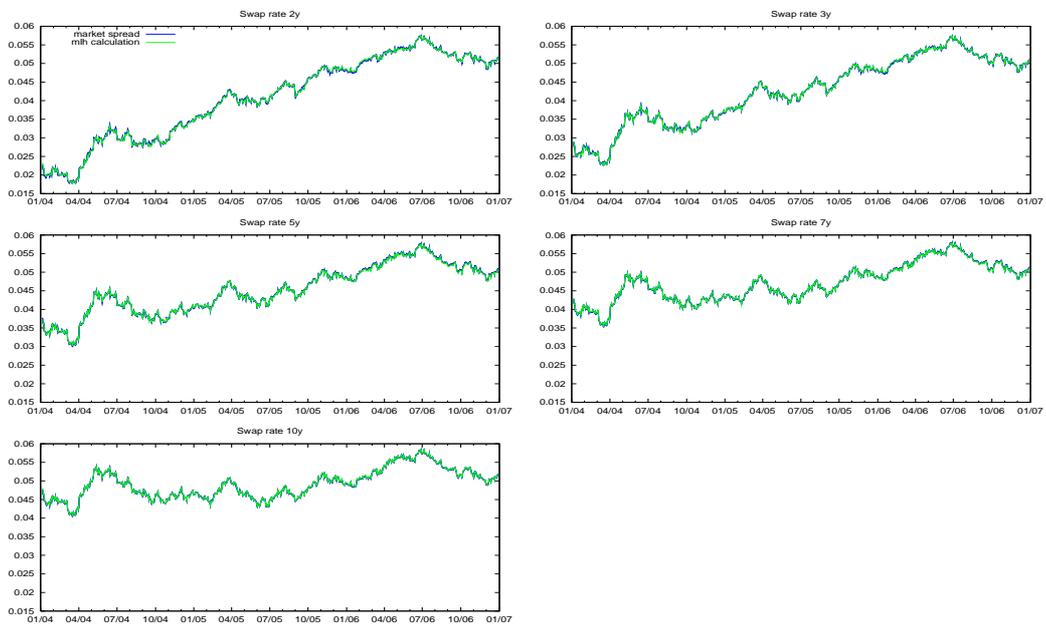}
  \caption{USD swap rates.}
  \label{prices_swap_usd}
\end{figure}

\begin{figure}
  \centering
  \includegraphics[width=0.95\textwidth]{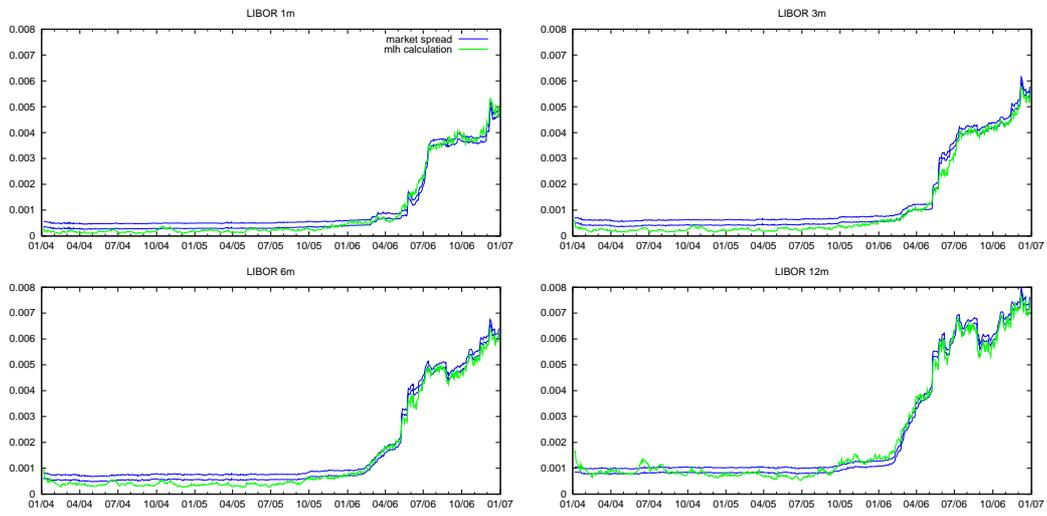}
  \caption{JPY Libor rates.}
  \label{prices_libor_jpy}
\end{figure}

\begin{figure}
  \centering
  \includegraphics[width=0.95\textwidth]{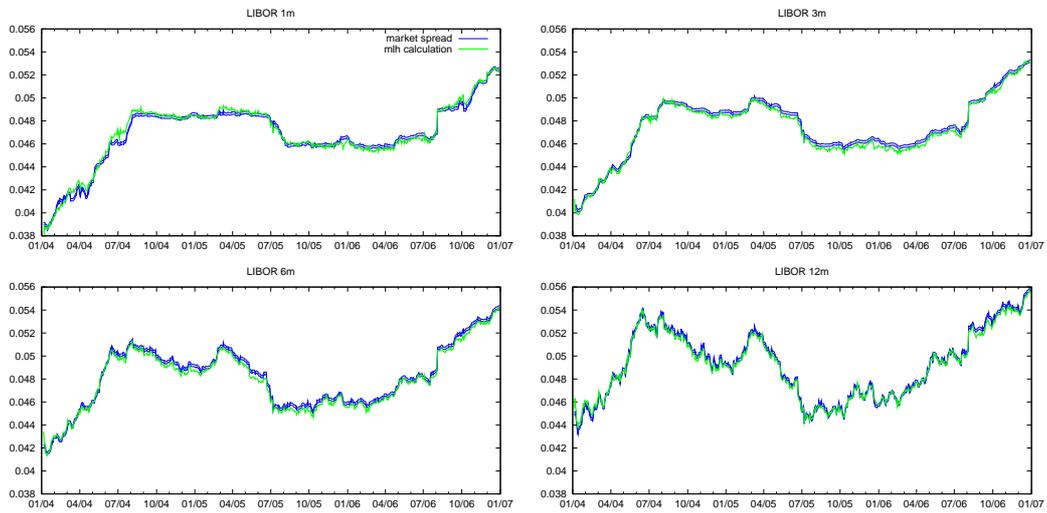}
  \caption{GBP Libor rates.}
  \label{prices_libor_gbp}
\end{figure}

\begin{figure}
  \centering
  \includegraphics[width=0.95\textwidth]{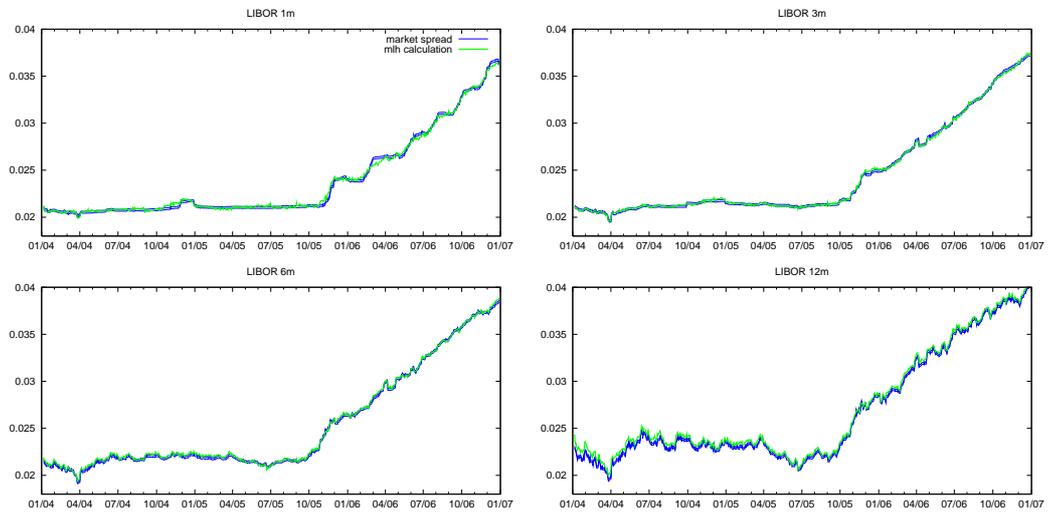}
  \caption{EUR Libor rates.}
  \label{prices_libor_eur}
\end{figure}

\begin{figure}
  \centering
  \includegraphics[width=0.95\textwidth]{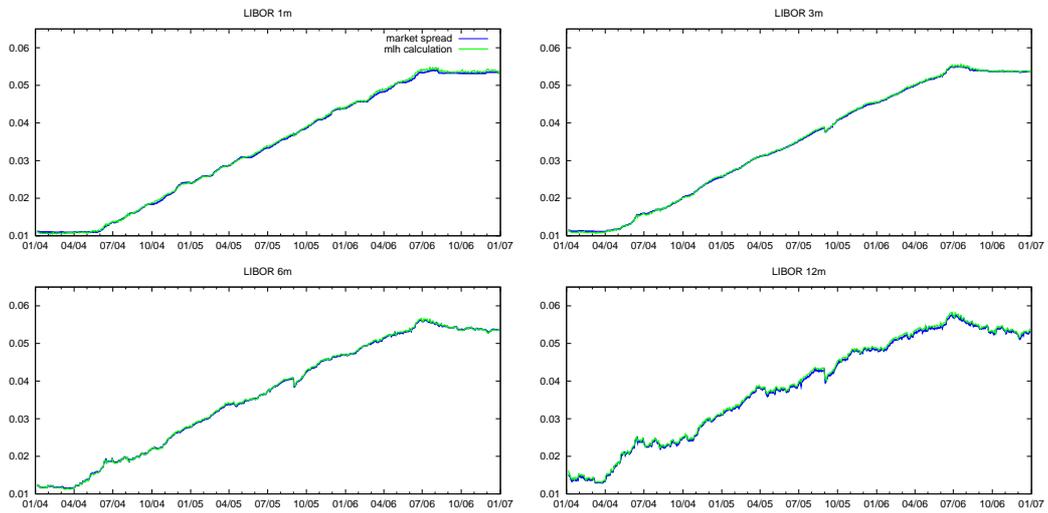}
  \caption{USD Libor rates.}
  \label{prices_libor_usd}
\end{figure}

There follow a number of plots of the ML fitted values\footnote{
The particle-filter population averages are generally quite close
to the ML values, and are omitted from these plots to aid clarity.
} (in green) and the 
corresponding market bids and asks (in blue) for various series:
FX forward rates (Figures \ref{prices_forward_jpy}, \ref{prices_forward_gbp},
\ref{prices_forward_eur});
swaption prices (Figures \ref{prices_swaption_jpy}, \ref{prices_swaption_gbp}, 
\ref{prices_swaption_eur},  \ref{prices_swaption_usd}); 
cap prices (Figures \ref{prices_cap_jpy}, \ref{prices_cap_gbp}, 
\ref{prices_cap_eur}, \ref{prices_cap_usd});
swap rates (Figures \ref{prices_swap_jpy},  \ref{prices_swap_gbp}, 
 \ref{prices_swap_eur},  \ref{prices_swap_usd});
Libor rates (Figures \ref{prices_libor_jpy}, \ref{prices_libor_gbp}, 
\ref{prices_libor_eur}, \ref{prices_libor_usd}).
The quality of the fits is visible from these plots. Perhaps the caps
work least well, with the swaptions also less good than the 
FX forwards, the swap reated and the Libor rates. It is perhaps not
too surprising that the OTC derivatives are less well fitted than the 
more liquid fundamentals, but this does highlight an area for further
work. For example, we have only reported on the fits obtained with
nearest-neighbour Markov chains on a circular state-space, and 
with never more than 7 states. There is therefore scope to improve
the fit by relaxing these restrictions, but the increased dimensionality
that would follow may mean that the fit is not much improved, if
at all.  We hope to be able to follow this further in subsequent work.

\section{Hedging.}\label{S5}
In conventional models, the standard way to hedge a
derivative is to {\em delta-hedge} it. We
compute the differential of the price of the derivative with
respect to the prices of the underlying instruments, and this
tells us how many units of the
underlying to hold to protect (to leading order) against
the moves in the underlying. In the case of a complete
market, this hedging methodology  perfectly replicates
the contingent claim we were trying to hedge.

If we are using a Markov chain potential model, the notion of
differentiating has no meaning, nevertheless the idea of
immunising our portfolio against possible changes will work just as
well.  Suppose that we have a derivative $Z$, and hedging instruments
$z^{(1)},z^{(2)},\ldots$. Suppose that if the state of
the chain at time $t$ is $i$ and it jumps to $j$ then the value of $Z$ changes
by $\Delta Z_{ij}(t)$. Then what we will do is to hold $w_r(t)$
units of asset $r$ so that
\begin{equation}
  \Delta Z_{ij}(t) + \sum_{r=1}^m \; w_r(t) \Delta z^{(r)}_{ij}(t) = 0
\quad \forall j  \quad(X_t =i).
\label{hedge1}
\end{equation}
Thus whatever jumps of the chain occur, our hedging portfolio
will be unaffected by them.  Of course, we do not in practice
 know $X_t$, but this does not alter the hedging
methodology; we would now make a portfolio of more hedging assets
so as to ensure that
\begin{equation}
  \Delta Z_{ij}(t) + \sum_{r=1}^M \; w_r(t) \Delta z^{(r)}_{ij}(t) = 0
\quad \forall i,j.
\label{hedge2}
\end{equation}
Following this recipe in the case of (say) a 5-state chain would
entail taking a position in 20 different hedging instruments (if
that many were available!) 

In the context of the particle-filtering modelling, the simplest thing we
could propose is to calculate the hedging requirement
for each particle in the population using the analysis of (\ref{hedge1})
 above (recall that each particle thinks it knows for certain what the 
state of the chain is). Taking a weighted average of the individual
particles' hedging requirements then gives a  first candidate for the hedge.

\begin{figure}
  \centering
  \includegraphics[width=0.95\textwidth]{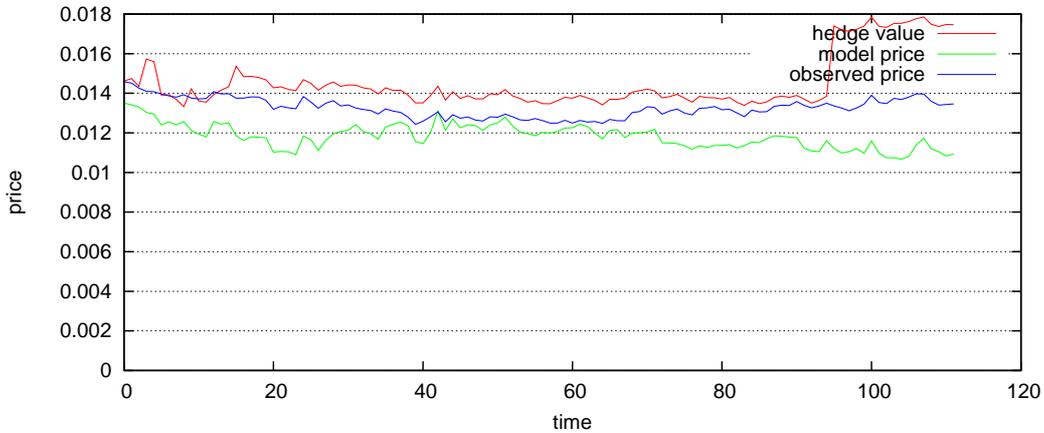}
  \caption{Here we hedge a swaption 5 into 2 years by caps:
	1, 3, 5, 7, 10 years in a 5 state chain. 
}
\label{plot5}
\end{figure}

In Figure \ref{plot5} we see how this simple-minded procedure performed
when we tried to hedge a 5-into-2 year swaption using some caps 
 The hedge is fitting the market price generally
as well as the model, and also appears to be tracking the underlying
very well; rises and falls in the underlying are accompanied by 
corresponding rises and falls in the value of the hedge.

\section{Conclusions.}\label{S6}
The calibration study conducted in this paper is a major and 
ambitious test of the concept that the potential approach may 
account {\em simultaneously} for the prices of many assets
in different currencies. Altogether, across the four currencies
considered, we were computing simultaneously the prices of 
168 instruments, and after time for the particle-filtering algorithm
to bed in, we were able to fit market prices to within an average
error of a  few 
spreads for all the instruments, sometimes much better.  The 
simple-minded hedging rules suggested by the modelling approach
gave hedge values which were quite close to the underlying, and
tracked well in the sense that the increments processes looked
quite similar.  

There remain further challenges to tackle, particularly in extending
the calibration to other classes of assets. The extension to credit
derivatives is mathematically relatively straightforward; the default of 
a firm is modelled by a credit spread which depends on the state of 
the Markov chain $\xi$, and this then needs to be estimated. What is
easy about this is that the pricing of CDOs and CDS is mathematically
very similar to pricing of riskless interest-rate derivatives. What is 
less appealing is the feature that one may need in principle a
different credit spread function for each firm, hugely increasing
the dimension of the problem. We expect that the correct approach
to this is to firstly fit and fix the model for riskless interest rates, and
then calibrate  firm- or sector-specific default intensities thereafter.

The next more challenging issue is to try to fit equities to the modelling
framework. At one level, this need not be
so hard, if we model the price of a stock as the NPV of all future dividends,
and then try to write the dividend process as a function of the underlying
Markov chain $\xi$. This introduces (in principle) a separate function
for each stock being considered, and again the approach will be to fit
the model to the big fixed-income, futures, FX data, then try to fit the
individual stock characterstics into that model.  However, it may turn 
out to be necessary to introduce individual Brownian terms into the 
individual stock. At very least, some translation of the proposed (discrete
state space) model into the more familiar terms of growth rate and
volatility will be necessary.

These are issues which remain to be tackled, and we hope that these
will be dealt with shortly. However, what is clear is that the Markov
chain potential approach which we advocate in the study has shown
an amazing capacity provide a model which closely fits major
fixed-income and FX assets in multiple currencies. This is important
at various levels, not least that it offers a framework for the pricing 
and hedging of hybrid derivatives of arbitrary complexity. The fitted
model  is not merely a fit; it
makes predictions about the co-movement of many assets, and so could
for example be used to price quite complicated credit derivatives
(a theme developed in the study \cite{DGR}).

%
%


\pagebreak

\pagebreak
\pagebreak
\bibliography{pfpot}

\end{document}